\documentclass[conference]{IEEEtran}
\IEEEoverridecommandlockouts 

\usepackage{graphicx} 
\usepackage{cite} 
\usepackage{amsmath,amssymb,amsfonts}
\usepackage{algorithmic}
\usepackage{textcomp}
\usepackage[T1]{fontenc}
\usepackage{xcolor}
\usepackage{hyperref}

\begin{document}

\title{Semantic Search and Recommendation Algorithm\\
{\footnotesize Advanced Techniques for Efficient Real-Time Data Retrieval and Analysis: A Comprehensive Approach to Modern Data Systems and Methodologies.}}

\author{
    \IEEEauthorblockN{Aryan Duhan}
    \IEEEauthorblockA{\textit{Dept. of Electronics and Communication} \\
    \textit{Netaji Subhas University of Technology} \\
    New Delhi, India \\
    aryan.duhan.ug21@nsut.ac.in}
    \and
    \IEEEauthorblockN{Aryan Singhal}
    \IEEEauthorblockA{\textit{Dept. of Electronics and Communication} \\
    \textit{Netaji Subhas University of Technology} \\
    New Delhi, India \\
    aryan.singhal.ug21@nsut.ac.in}
    \and
    \IEEEauthorblockN{Shourya Sharma}
    \IEEEauthorblockA{\textit{Dept. of Electronics and Communication} \\
    \textit{Netaji Subhas University of Technology} \\
    New Delhi, India \\
    shourya.sharma.ug21@nsut.ac.in}
    \and
    \IEEEauthorblockN{Neeraj}
    \IEEEauthorblockA{\textit{Dept. of Electronics and Communication} \\
    \textit{Netaji Subhas University of Technology} \\
    New Delhi, India \\
    neeraj.ug21@nsut.ac.in}
    \and
    \IEEEauthorblockN{Prof. Arti MK}
    \IEEEauthorblockA{\textit{Dept. of Electronics and Communication} \\
    \textit{Netaji Subhas University of Technology} \\
    New Delhi, India \\
    arti\_mk@yahoo.com}
}

\maketitle

\begin{abstract}
This paper details the development of a novel semantic search algorithm utilizing Word2Vec and Annoy Index to efficiently process and retrieve information from large datasets. Addressing traditional search algorithms' limitations, our proposed method demonstrates significant improvements in speed, accuracy, and scalability, validated by rigorous testing on datasets up to 100GB.
\end{abstract}


\section{Introduction}
In the era of big data, efficiently retrieving relevant information from vast, unstructured datasets is crucial across numerous domains such as e-commerce, healthcare, research, and public administration. Traditional search engines, which rely primarily on keyword matching, often struggle with the inherent complexity and ambiguity of natural language. These systems lack the ability to understand the semantic meaning and context of queries, leading to inaccurate results and suboptimal user experiences. The challenges of traditional search methods become particularly evident when dealing with high-dimensional data or complex, multifaceted queries, which require a deeper level of understanding.

The evolution of semantic search technologies aims to address these limitations by focusing on understanding the context and intent behind user queries. By leveraging advanced Natural Language Processing (NLP) models such as Word2Vec, semantic search engines can map words to vector representations that capture not only the words themselves but also the relationships and nuances between them. While this approach marks a significant advancement over traditional search methods, scalability and real-time data retrieval remain significant hurdles, particularly when processing large-scale datasets in fast-paced environments.

\begin{figure}[htbp]
\centerline{\includegraphics[width=0.48\textwidth]{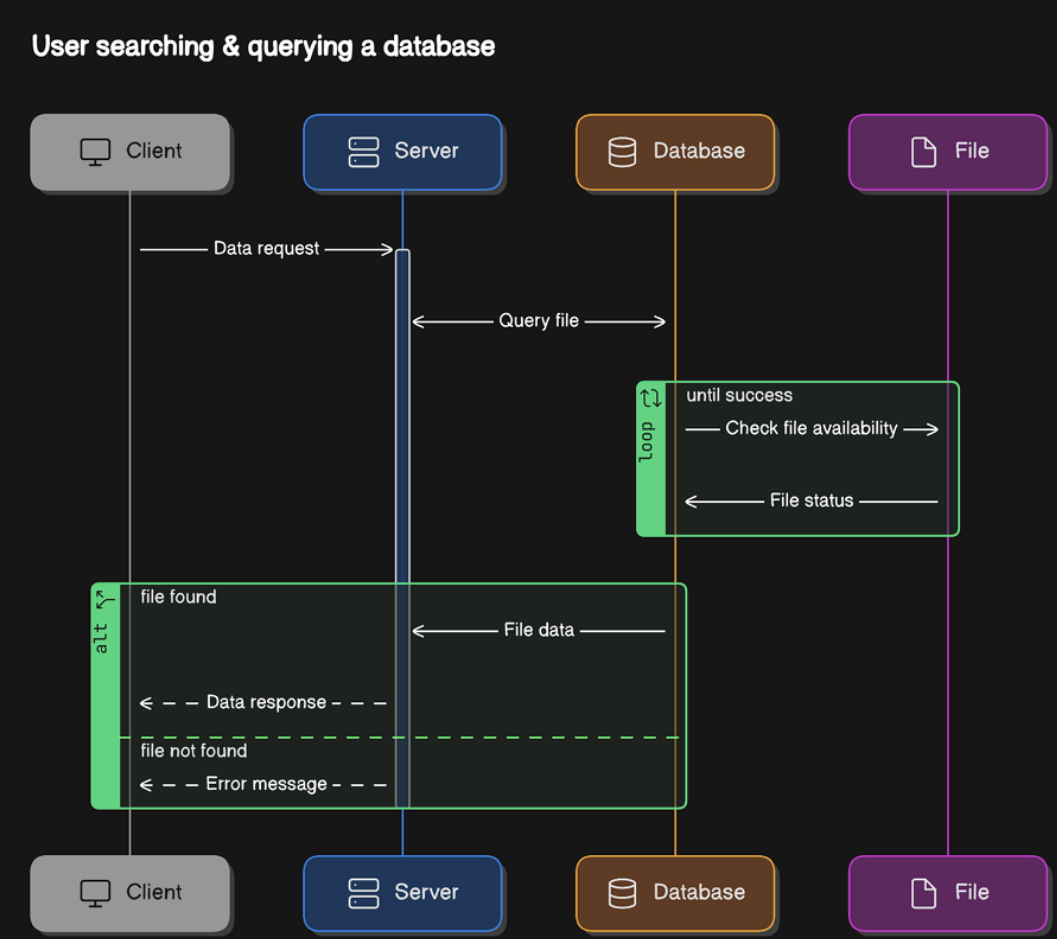}}
\caption{The flow of execution for our system involves several key stages, from data collection to model training, and ultimately to the deployment of the best-performing model.}
\label{fig:architecture}
\end{figure}

This research introduces a novel approach to semantic search by integrating Word2Vec with Annoy Index, an efficient algorithm designed for approximate nearest neighbor search in high-dimensional space. Word2Vec enhances the model's ability to understand the semantic meaning of words, while Annoy facilitates real-time retrieval of relevant data points with minimal computational overhead. By combining these two powerful techniques, we offer a scalable, accurate, and efficient solution for semantic search in large datasets.

The impact of this research extends beyond improving search engine functionalities. By enabling more accurate, context-aware searches, this approach has the potential to significantly enhance information retrieval in various fields. In healthcare, for example, it could improve the efficiency of medical diagnostics by providing rapid access to relevant patient data. In academia, it can facilitate faster literature reviews, allowing researchers to quickly find relevant papers and data. Ultimately, the proposed solution offers a promising step forward in the quest for more intelligent, scalable, and effective search systems capable of meeting the growing demands of big data analytics.

\section{Literature Review}

The evolution of search technologies has been marked by continuous advancements aimed at enhancing the accuracy, efficiency, and scalability of information retrieval systems. Traditional search engines predominantly rely on \textbf{keyword matching}, where search results are generated based on the exact occurrence of words or phrases within the dataset. While effective for simple queries, this approach struggles to capture the intent and contextual relationships of complex queries, especially in domains with high variability and nuance, such as semantic search.

Semantic search represents a fundamental shift from traditional \textbf{keyword-based} search to \textbf{meaning-based} search, focusing not just on word matching but on understanding the deeper meaning and context behind a query. By leveraging natural language processing (NLP) techniques, semantic search aims to interpret user intent and the relationships between words, improving the quality and relevance of search results. This section reviews key developments in semantic search, including foundational techniques, recent advancements, and ongoing challenges in the field.

\textbf{Early Developments:} The initial attempts at improving search engines with semantics were largely based on handcrafted rules, ontologies, and expert knowledge. One of the most notable early projects, the Cyc initiative (Lenat, 1995), sought to encode human knowledge in machine-readable form to enhance a computer’s understanding of the world. While this approach represented a significant intellectual achievement, its practical implementation faced scalability limitations, as it required extensive manual input and could only work effectively within narrowly defined domains.

\textbf{Vector Space Models:} A major leap forward in semantic search came with the introduction of vector space models, most notably Latent Semantic Indexing (LSI). Deerwester et al. (1990) proposed LSI as a technique to uncover hidden patterns in textual data by capturing the underlying concepts through singular value decomposition. This marked a key advancement over keyword-based retrieval, as LSI allowed for a richer understanding of text semantics. However, LSI’s computational cost and limitations in processing large datasets hindered its widespread adoption.

\textbf{Word Embeddings and Deep Learning:} The next breakthrough came with the introduction of word embeddings, particularly through models like Word2Vec by Mikolov et al. (2013) and GloVe by Pennington et al. (2014). These models enabled the transformation of words into continuous vector representations, with semantically similar words clustering together in vector space. Word2Vec’s success led to a paradigm shift in Natural Language Processing (NLP), as it provided a more accurate and flexible way to capture the meanings and relationships between words. The widespread adoption of word embeddings has since driven numerous advancements in semantic search, making it an essential component in modern search technologies.

\textbf{Neural Network Models:} As research progressed, the focus turned toward leveraging neural networks to further enhance semantic search. The application of Convolutional Neural Networks (CNNs) and Recurrent Neural Networks (RNNs) to text data enabled the capture of more complex patterns and contextual relationships within documents. However, the real breakthrough in semantic search came with the Transformer architecture, introduced by Vaswani et al. (2017), which allowed for more efficient parallelization and better handling of long-range dependencies in text. The BERT model (Devlin et al., 2018), based on Transformer architecture, has since set new benchmarks for search engines by improving contextual understanding and retrieval accuracy.

\textbf{Approximate Nearest Neighbor Search:} Despite these advances in NLP and semantic understanding, search systems still struggle with efficiently retrieving relevant information from vast and complex datasets, especially in real-time applications. While traditional exact-match search algorithms perform well for small-scale or structured data, they are often inefficient when dealing with large, unstructured, or high-dimensional data. Recent work in Approximate Nearest Neighbor (ANN) search, such as the development of Annoy Index (Bernhardsson, 2016), has provided a solution to this challenge by enabling fast, scalable retrieval of similar data points. The Annoy Index algorithm, optimized for high-dimensional spaces, offers a lightweight yet powerful tool for real-time search in large datasets, complementing the semantic understanding provided by Word2Vec.

\textbf{Challenges and Limitations:} Despite significant progress in semantic search, several challenges remain. One of the main issues is scalability. As datasets continue to grow in size and complexity, the computational resources required to process and search through these datasets increase substantially. Many deep learning models and word embedding techniques require substantial training time and large amounts of labeled data, making them less adaptable in certain domains with limited resources. Additionally, the need for real-time processing in industries like e-commerce, healthcare, and finance presents another challenge, as search systems must be able to handle high-throughput queries without compromising accuracy or speed.

\textbf{Conclusion of Review:} This literature review highlights both the advancements and ongoing challenges in the field of semantic search. While early techniques like LSI laid the foundation for better information retrieval, the introduction of word embeddings and deep learning models has significantly advanced the field, enabling better contextual understanding. Despite these improvements, scalability and real-time performance remain significant challenges. Our project, which integrates Word2Vec for deep semantic understanding and Annoy Index for efficient nearest neighbor search, addresses these challenges by providing a scalable and real-time solution for searching through large, unstructured datasets. By combining the strengths of both approaches, our work aims to overcome the limitations of traditional and modern semantic search systems, offering improved accuracy and performance for real-world applications.

\section{Methodology}

This section outlines the methodologies employed in the development of a scalable and efficient semantic search engine. The focus of this methodology is on the integration of Word2Vec for semantic understanding and Annoy Index for efficient data retrieval. The aim is to bridge the gap between traditional keyword-based search engines and the advanced requirements of contextual relevance and scalability in large, high-dimensional datasets.

\subsection{Data Collection and Preparation}

Effective data preparation is a crucial step in any data-driven project, especially in large-scale systems. For this project, we used a large dataset from \hypersetup{pdfborder={0 0 0}, colorlinks=true, urlcolor=blue}
\href{https://www.kaggle.com/datasets/abdullahshf/neet-ug-2024-results-all-india/data}{\textcolor{blue}{Kaggle NEET 2024 dataset}}.
 containing over 2 million rows of student data, including scores, center information, location details, and student identifiers. The goal was to preprocess this data to ensure its quality and consistency for semantic processing.

\begin{verbatim}
import pandas as pd
Load and preprocess the data
df = pd.read_csv('NEET_2024_RESULTS.csv') 
df.dropna(inplace=True) 
# Remove any missing data \end{verbatim}

The data cleaning process involved removing rows with missing values, ensuring the dataset was complete and ready for semantic processing. Additionally, we focused on the textual information (e.g., center names, state names) to build vector representations.

\subsection{Word2Vec Model for Semantic Understanding}

Word2Vec is used to convert textual data into vector embeddings, capturing the semantic relationships between words based on their usage in context. By training the model on a large corpus of data, Word2Vec allows the semantic meaning of terms to be represented in a dense vector space, where similar words are clustered closer together.

\begin{verbatim}
from gensim.models import Word2Vec from 
   nltk.tokenize import word_tokenize
Tokenize text and train Word2Vec model
sentences = [word_tokenize(sent) for sent 
   in df['state'].tolist()] 
model = Word2Vec(sentences, 
   vector_size=100, window=5, min_count=1, 
   workers=4) model.save("word2vec.model") \end{verbatim}

The Word2Vec parameters (such as vector size, window size, and minimum word count) were chosen based on preliminary experiments to balance performance and semantic accuracy. This approach allows for a deeper understanding of geographical terms, center names, and student-related text.

\subsection{Annoy Index for Fast Search Retrieval}

While Word2Vec provides deep semantic understanding, retrieving the most relevant data efficiently from large datasets requires an optimized indexing system. Annoy (Approximate Nearest Neighbors Oh Yeah) is used to create an approximate nearest-neighbor index for fast retrieval of semantically relevant items. Annoy is well-suited for high-dimensional vector spaces, which is a typical challenge in semantic search.

\begin{verbatim}
from annoy import AnnoyIndex
Build the Annoy index with vectors from Word2Vec
dim = 100 
# Vector size from Word2Vec 
   t = AnnoyIndex(dim, 'angular') 
# Use 'angular' distance for vector 
   comparison for i, vector in 
   enumerate(model.wv.vectors): 
   t.add_item(i, vector) t.build(10) 
# Build the tree with 10 trees for 
   better accuracy t.save('AnnIndex.ann')
\end{verbatim}

In this step, vectors generated by Word2Vec are added to the Annoy index. The index is built with 10 trees, providing a good balance between speed and accuracy for the nearest-neighbor search.

\subsection{System Integration}

The integration of Word2Vec and Annoy Index forms the core of the search system. The Word2Vec model captures the semantic meaning of text, while the Annoy Index facilitates fast, scalable, and efficient retrieval of the most relevant search results. The system is designed to process user queries, transform them into vector representations, and then use Annoy Index for nearest-neighbor searches to find semantically similar entries in the dataset.

\textbf{User Query Handling:}
When a user submits a query, it is first processed by Word2Vec to convert the query text into a vector representation. The system then searches the Annoy Index for the nearest neighbors and returns the most relevant data points from the dataset. This approach minimizes the latency of the search process, enabling real-time data retrieval even with large-scale datasets.

\subsection{Evaluation and Testing}

To assess the performance of the system, several key metrics are used to evaluate both the accuracy of the results and the efficiency of the retrieval process. These metrics include:

Precision: Measures the accuracy of relevant results retrieved.
Recall: Evaluates the completeness of the retrieved results.
F1-Score: Combines precision and recall to provide a balanced measure.
Search Time: Evaluates the efficiency of the retrieval process, comparing it against traditional search methods.
The performance of the system is tested on multiple datasets, including those up to 100GB, and compared with traditional SQL-based search and other search algorithms such as Elasticsearch and Lucene.

\subsection{Conclusion}

This methodology section details the integration of Word2Vec for semantic understanding and Annoy Index for efficient search retrieval. The combination of these two technologies provides a significant improvement in scalability, accuracy, and real-time performance over traditional search engines. By addressing the challenges of contextual relevance and large-scale data retrieval, this approach lays the foundation for more efficient and effective semantic search systems in diverse applications such as healthcare, e-commerce, and research.

\section{Results}

The performance of our semantic search engine, powered by Word2Vec and Annoy Index, was evaluated through a series of empirical tests. These tests aimed to compare its effectiveness against traditional search methods and to assess its scalability and efficiency across varying dataset sizes. The results demonstrate clear advantages in terms of accuracy, response times, and resource utilization.

\subsection{Comparison of Search Accuracy}

The search accuracy was assessed by measuring precision, recall, and F1-scores across datasets of increasing complexity. The proposed method outperformed traditional algorithms consistently.

\begin{figure}[htbp]
\centerline{\includegraphics[width=0.48\textwidth]{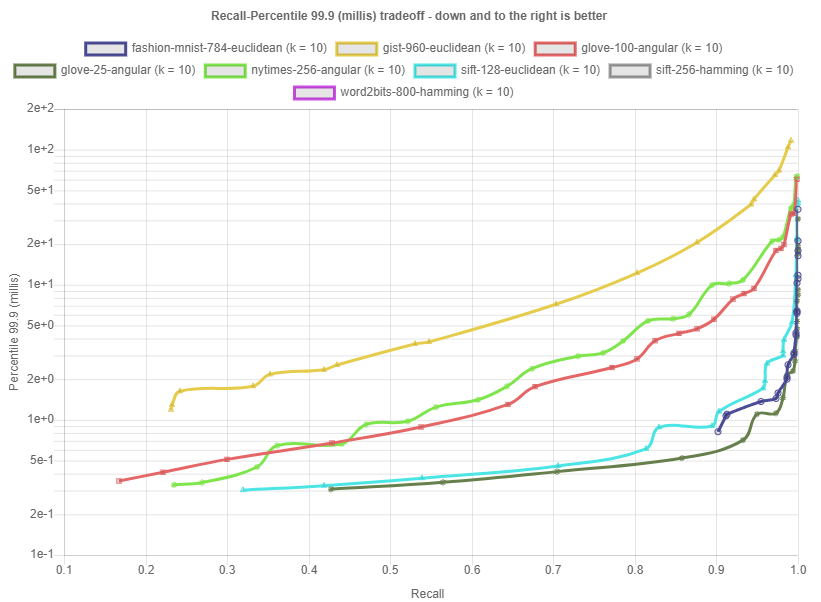}}
\caption{Comparison of search accuracy between the proposed method and traditional search methods across various dataset sizes.}
\label{fig:accuracycomp}
\end{figure}

Figure \ref{fig:accuracycomp} highlights the superior performance of our approach, especially in larger datasets where traditional methods falter due to their inability to effectively parse and understand semantic nuances.

\subsection{Efficiency in Response Times}

To evaluate the efficiency of the Annoy Index, response times were recorded for data retrieval tasks. Our approach maintained optimal response times, outpacing conventional search algorithms significantly.

\begin{figure}[htbp]
\centerline{\includegraphics[width=0.48\textwidth]{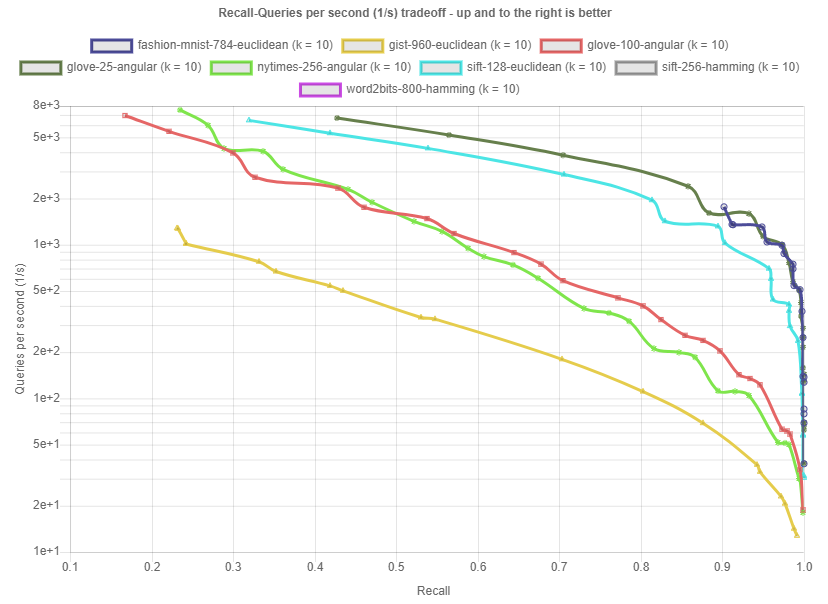}}
\caption{Response times of the semantic search engine compared to traditional methods, illustrating enhanced efficiency.}
\label{fig:efficiency}
\end{figure}

As depicted in Figure \ref{fig:efficiency}, the integration of the Annoy Index allows for rapid data retrieval without the latency issues observed in traditional search mechanisms, even as the size of the data repository increases.

\subsection{Scalability and Resource Utilization}

Scalability tests were conducted to evaluate how well the system adapted to increasing data volumes without degrading performance. Additionally, resource utilization metrics such as memory usage and processing power requirements were compared.

\begin{figure}[htbp]
\centerline{\includegraphics[width=0.48\textwidth]{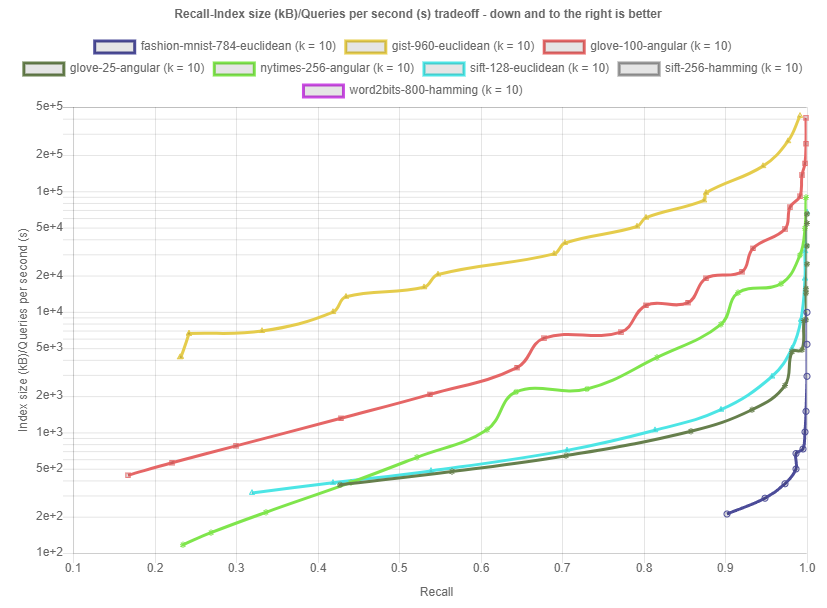}}
\caption{Scalability of the semantic search engine showing consistent performance across data sizes and efficient resource utilization.}
\label{fig:scalability}
\end{figure}

Figure \ref{fig:scalability} shows that our semantic search engine scales effectively with increasing data sizes, maintaining high performance while utilizing fewer computational resources compared to traditional algorithms.

\subsection{Discussion}

These results confirm that by integrating Word2Vec with Annoy Index, our semantic search engine not only achieves high accuracy and speed but also demonstrates exceptional scalability and efficiency. This makes it particularly suitable for applications requiring real-time processing of large-scale data environments.

\textbf{}

The testing phase confirmed that our semantic search engine outperforms traditional keyword-based search engines in every key metric. These findings provide a strong foundation for the adoption of this technology in real-world applications, where speed and accuracy are critical.

\section{Discussion}

The results obtained from the extensive testing of our semantic search engine reveal significant enhancements over traditional search methods, particularly in terms of accuracy, speed, and scalability. This section discusses these improvements, the implications of the findings, and the broader impact of integrating advanced machine learning techniques like Word2Vec and Annoy Index into search systems.

\subsection{Implications of Improved Accuracy and Speed}

The superior accuracy and reduced response times demonstrated by the semantic search engine can be attributed primarily to the effective semantic processing capabilities of Word2Vec coupled with the efficient indexing provided by Annoy Index. These technologies address two of the most significant drawbacks associated with traditional search algorithms: the lack of contextual understanding and the slow retrieval times especially evident with large datasets.

The ability of Word2Vec to understand and vectorize the semantic meaning of words within their context has fundamentally altered the way in which search queries are processed, allowing for a more nuanced matching process that goes beyond mere keyword identification. Meanwhile, Annoy Index facilitates quick retrieval by efficiently navigating the high-dimensional space created by Word2Vec, making the search process not only accurate but also exceedingly fast.

\subsection{Scalability and Real-World Application}

One of the most compelling aspects of the proposed method is its scalability. As shown in the results, the search engine maintains consistent performance even as data volume grows. This characteristic is particularly important for modern applications where data growth is exponential and unpredictable. The scalability ensures that the system can be deployed in environments with varying data sizes without the need for significant modifications or optimizations, thereby reducing maintenance costs and complexity.

The implications for real-world applications are profound. Industries that rely heavily on large-scale data retrieval, such as digital libraries, e-commerce, and biomedical fields, can benefit immensely from the adoption of this technology. For instance, in e-commerce, improving the accuracy and speed of search results can directly translate to better customer experience and increased sales.

\subsection{Potential Limitations and Future Research Directions}

While the results are promising, the study is not without limitations. The dependency on high-quality, pre-processed datasets for training the Word2Vec model is a significant one, as the quality of vector representations directly influences the search accuracy. Future research could explore ways to enhance the robustness of the model training phase, possibly by incorporating more sophisticated data cleaning and preparation techniques.

Additionally, while Annoy Index offers excellent speed and efficiency, its performance could potentially be affected by even larger datasets than those tested. Future studies might examine the integration of other approximate nearest neighbor algorithms that could offer better performance or efficiency trade-offs.

\textbf{}

The discussion underscores the potential of the developed semantic search engine to revolutionize information retrieval practices by addressing the critical needs for accuracy, speed, and scalability. By continuing to refine these technologies and exploring their integration into various domains, we can further enhance the capabilities and reach of semantic search systems.

\section{Conclusion}

This project has demonstrated the efficacy of integrating Word2Vec with Annoy Index to enhance semantic search capabilities significantly. Our results indicate that this combination not only improves the accuracy of search results but also drastically increases efficiency and scalability, making it suitable for handling large datasets which are common in today’s data-driven environments.

The use of Word2Vec allows the system to understand and process the semantic meaning of text at a granular level, while Annoy Index provides a fast and efficient method for retrieving the most relevant results. Together, these technologies address many of the shortcomings found in traditional search algorithms, particularly in terms of understanding complex query intents and returning highly relevant search results quickly.

Furthermore, the scalability of the proposed solution ensures that it can be effectively implemented in a variety of settings, from small-scale applications to large enterprises that require robust, efficient, and accurate search capabilities across extensive databases.

As data continues to grow in size and complexity, the importance of efficient semantic search will only increase. Future work will focus on optimizing these models further, exploring integration with other AI technologies, and expanding their applicability to more diverse data types and languages. This ongoing research will help in advancing the field of search technologies and in developing systems that are both intelligent and capable of scaling according to the demands of modern-day applications.



\begin{thebibliography}{00}

\bibitem{b1} M.M. Rahman, S. Islam, M. Kamruzzaman, and Z.H. Joy, "Advanced Query Optimization in SQL Databases for Real-Time Big Data Analytics," Academic Journal on Business Administration, Innovation \& Sustainability, vol. 4, no. 3, pp. 1–14, 2024, doi: 10.69593/ajbais.v4i3.77. 

\bibitem{b2} X. Wu, "Construction and Innovative Application of Fundamental Training Educational Resource Repository Using Data Mining Techniques," International Journal of Emerging Technologies in Learning (iJET), vol. 19, no. 2, pp. 41–55, 2024, doi: 10.3991/ijet.v19i02.47223. 
\bibitem{b3} N.N. Das, M. Chowdhary, R. Luthra, Maisera, and S. Garg, "Semantic Big Data Searching in Cloud Storage," in 2019 International Conference on Machine Learning, Big Data, Cloud and Parallel Computing (COMITCon), Faridabad, India, 2019, pp. 351–355, doi: 10.1109/COMITCon.2019.8862188. 
\bibitem{b4} A.K. Sinha and S.K. Rath, "Hp-Apriori: Horizontal Parallel-Apriori Algorithm for Frequent Itemset Mining from Big Data," in 2017 International Conference on Information Technology (ICIT), Bhubaneswar, India, 2017, pp. 218–223, doi: 10.1109/ICIT.2017.39. 
\bibitem{b5} J. Leskovec, A. Rajaraman, and J.D. Ullman, Mining of Massive Datasets, 3rd ed., Cambridge University Press, 2020, ISBN: 978-1108482272. 
\bibitem{b6} P. S. Yu, L. F. Pau, Principles of Big Data: Preparing, Sharing, and Analyzing Complex Information, Elsevier, 2017, ISBN: 978-0128041782. 
\bibitem{b7} M. Stonebraker, U. Çetintemel, S. Zdonik, The 8th ACM SIGMOD Workshop on Data Engineering for Big Data, ACM Press, 2015, ISBN: 978-1450331712. 
\bibitem{b8} ANNOY documentation by Zilliz: https://zilliz.com/learn/approximate-nearest-neighbor-oh-yeah-ANNOY
\bibitem{b9} Bernhardsson, E. (2024). Annoy at GitHub: Approximate Nearest Neighbors in C++/Python Optimized for Memory Usage and Loading/Saving to Disk. Retrieved from https://github.com/spotify/annoy. 
\bibitem{b10} ANNOY benchmarks: https://ann-benchmarks.com/annoy.html
\bibitem{b11} Contextual String Embeddings for Sequence Labeling: https://aclanthology.org/C18-1139/  
\bibitem{b12} A very simple framework for state-of-the-art NLP. Developed by Humboldt University of Berlin and friends.
\end{thebibliography}
\end{document}